\documentstyle[12pt]{article}

\font\elevenbf=cmbx10 scaled\magstep 1
 1
    \newcommand{\r}[1]{(\ref{#1})}
    \newcommand{\beq}{\begin{equation}}
    \newcommand{\eeq}{\end{equation}}
    \newcommand{\EQN}{\label}
\renewcommand{\arraystretch}{2}                             %
 \newcommand{\as}{\alpha_s}

\newcommand{\ac}{a}
\newcommand{\Ac}{a}

\newcommand{\ovl}{\overline}
\newcommand{\ba}{\begin{array}}
\newcommand{\ea}{\end{array}}
\newcommand{\al}{\alpha}

\newcommand{\be}{\beta}
\newcommand{\g}{\gamma}
\newcommand{\G}{\Gamma}
\newcommand{\dsp}{\displaystyle}
\newcommand{\msbar}{\overline{\mbox{MS}}}

\newcommand{\D}{\Delta}
\newcommand{\gm}{\gamma_m}
\def\bbuildrel#1_#2^#3%
{\mathrel{\mathop{\kern 0pt#1}\limits_{#2}^{#3}}}

\newcommand{\trpsbs}[2]{_{{}_{\dsp #1_{{}_{\scriptstyle #2}}}}}
\newcommand{\ice}[1]{\relax}
\newcommand{\ep}{\epsilon}
\begin{document}
%%%%%%%%%%%%%%%%%%%%%%%%%%%%%%%%%%%%%%%%%%%%%%%%%%%%%%%%%%%%%

\begin{titlepage}
\noindent
\hfill TTP93--38\\
\mbox{}
\hfill INR-842/93\\
\mbox{}
\hfill  December 1993   \\
%\hfill \today \\
%
% Title
%
\vspace{0.5cm}
\begin{center}
%  \begin{Large}
  \begin{bf}
The $\alpha_s^3$   Corrections
to the Effective Neutral Current
and to  the Z Decay Rate
\\
in the Heavy Top Quark Limit
 \\
\end{bf}
%\end{Large}

\vspace{0.8cm}
%\begin{large}
K.G.Chetyrkin\footnote{On leave from Institute for Nuclear Research
of the Russian Academy of Sciences, Moscow, 117312, Russia.}
\ice{J.H.K\"uhn} \\[2mm]
    Institut f\"ur Theoretische Teilchenphysik\\
    Universit\"at Karlsruhe\\
    D-76128 Karlsruhe, Germany\\[2mm]
O.V.Tarasov\footnote{On leave from the Joint Institute
                     for Nuclear Research
                     Moscow, 141980, Dubna,  Russia.}
\\[2mm]
    Fakult\"at f\"ur  Physik, Universit\"at Bielefeld \\[1mm]
    D-33615 Bielefeld, Germany
%\end{large}
%
% Abstract
%

\vspace{1cm}
  {\bf Abstract}
\end{center}

\begin{quotation}
\noindent
We find the effective neutral current in the heavy top quark limit up
to and including the terms of order $\alpha_s^3$. The result is then
employed to compute the correction of the same order to the
axial  part of   the Z decay rate  into hadrons  ($\G^h_Z$)  induced
by the top-bottom quark mass splitting. The calculations confirm
the prediction made in a previous work of  K.G.Ch. and J.H.K\"uhn
(Phys. Lett. B 308 (1993) 127),
namely, the effect of the singlet corrections of order $\alpha^3_s$
to the axial part of the $\G^h_Z$ should not exceed 25\% - 30\%
(depending on the values chosen for $\alpha_s(M_Z)$ and $m_t$) of the
magnitude of the leading (singlet) $\alpha_s^2$ term.

\end{quotation}

\end{titlepage}

\noindent
{\large\bf 1. Introduction}
\vspace{4mm}

\noindent
The QCD corrections to
the hadronic decay rate of the Z boson ($\G^h_Z$) have been under
continuous investigation since it became clear that the high
precision LEP experiments \cite{LEP} make an accuracy well below the
percent level for theoretical predictions  mandatory.

The  decay rate
$\G_Z^h$  including  all strong interaction corrections
may be viewed as an incoherent sum of  vector
($\G^V_Z$) and axial ($\G^A_Z$)  contributions.
Up to order  of  $O(\al_s)$       and for massless
quarks the QCD corrections to  $\G^h_Z$ are in
one-to-one correspondence  to those for
$R(s) =
\sigma_{tot}(e^+ \, e^- \to \mbox{{\rm h}})/
\sigma (\mu^+ \, \mu^-  \to e^+ \, e^-).
$
This is the case because for massless outgoing quarks the relevant
two-point functions are simply related
(see below) to the electromagnetic current
two-point function including terms of order $\alpha_s$.
Moreover, for the vector  part $\G_Z^V$ the  one-to-one diagram-wise
correspondence takes place in all orders of
perturbation theory. The latter
allows to use the results of \cite{Sapirstein,Gorishny1} to deduce
as many as three successive $\al_s$ corrections
to   $\G_Z^V$ (see  e.g.  a recent review \cite{WH} and Section 4).
It is clearly   desirable to know also the
strong interaction corrections to $\G_Z^A$ with the
same accuracy. In this case however the top quark contributions
must be taken into account.

Being quite  heavy, the top quark brings  a new dimension to the
game. This is connected with the fact that within the standard model
one may not in general just  "decouple"  the quark from processes
taking place on energy scales less than the top quark mass
\cite{Veltman77,Sirlin80,Collins78}.
It was first shown in works
\cite{Kniehl90a,Kniehl90b} by means of a direct completely analytical
calculation that  in the  $\alpha^2_s$ order   there appears a
new contribution to  $\G_Z^A$
peculiar to the axial vector current correlator.
Their result for
the partial decay rate of the $Z$ boson to $b$ quarks and gluons
$\G^A(Z\to b \ovl b)$  is of the form
(in the following we use the
effective couplant $a(\mu) \equiv \frac{\dsp\as(\mu)}{\dsp\pi}$)
\beq
\ba{c} \G^A(Z\to b \ovl b)=\G^{QPM} \left[
 1 + a(\mu) + a^2(\mu)(1.409  +
I_2(r)/3 )                    \right],
 \ \
\G^{QPM}=\frac{G_F M^3_Z}{8\pi\sqrt2},
\\ \frac{1}{3} I_2(r) =  \ln(4r) - 37/12 +(28/81)r +
0.211 \, r^2 + O(r^3),
\ \ r\equiv M_Z^2/(4m_t^2).
\ea
\EQN{Kuhn1}
\eeq
\ice{
Eq. \r{Kuhn1} clearly demonstrates that the top quark contribution
can not be naively ``decoupled" as $m_t \to \infty$.  }
Eq. \r{Kuhn1}
shows that any calculation of  the $O(\al_s^3)$ correction to the
$\G_Z^A$ must deal with   highly non-trivial {\em massive} four-loop
diagrams \ice{ (some examples are given in Fig.1 b,c) } -- a rather
hopeless task if one would try to get a complete answer for arbitrary
value of the ratio $r=M_Z^2/4m_t^2$  as was done in
\cite{Kniehl90a,Kniehl90b} for three loop diagrams. In addition,  the
very appearance of  potentially large logarithms of $r$ makes
it necessary to find a way for their resummation.

In \cite{CK3,CK4} it was  shown   that the standard methods
of  heavy mass expansion  with respect to the mass of the top
quark   or equivalently those of the effective field
theory \cite{Smi91} not only lead to a natural resummation  of
series of powers of $(\alpha_s \ln(M^2_Z/4m^2_t))$ appearing in higher
order corrections to \r{Kuhn1} but also  provide a  practical
calculational scheme to  proceed to higher orders.  In particular,
the first three terms  in the above formula for $d_2^{A,S}$ were
re-derived in a systematical manner  \cite{CK4,CK5} and also all
leading log and next to leading log QCD corrections of orders
$\as^{1+n}(\ln(M_Z^2/m^2_t))^n$ and $\as^{2+n}(\ln(M^2_Z/m^2_t))^n$
respectively were evaluated explicitly and summed up
\cite{CK4}. Exploiting  the
(non-physical) renormalization scale dependence of the
result  it was concluded in \cite{CK4} that the effect of the singlet
corrections of order $\alpha^3_s$ to the axial part of the $\G^h_Z$
might hardly be larger than $\pm$ 30\% of the
leading singlet $\alpha_s^2$ term.

In the present work we use the formalism  developed in \cite{CK3,CK4}
to  evaluate the effective neutral current  and  the axial part
of the Z boson hadronic decay rate in the heavy top quark limit up to
and including the terms of order $\alpha_s^3$.
Our calculations  confirm the predictions discussed above and
lead directly to the  complete  theoretical result for the
total decay rate of the Z boson with the $O(\al^3_s)$ accuracy
(see formulas (\ref{5.1},\ref{5.2})  in Section 5).

The paper is organized in the following way.  In Section 2 we
describe the calculation of the $O(\al^3_s)$ correction to the
effective topless neutral current which appears in the place of the
neutral current after the top quark is ``integrated out".
Section 3 deals with  the calculation of the $\al^3_s$
correction to the singlet part of the axial vector current correlator
in the massless case --- a necessary prerequisite to finding the
$\al^3_s$ correction to $\G_Z^A$.
In Section 4  we combine the results of the two previous sections to
produce the correction of order $\al^3_s$ to     $\G_Z^A$.
Our results are  discussed in Section 5.
Throughout the work  we  employ
dimensional regularization \cite{Hooft72,dim.reg}
and  the  $\msbar$ renormalization scheme \cite{ms,MSbar}.
All the calculations were performed  in a general
covariant gauge; the independence of the results from
the gauge fixing parameter  provides a  good  test
of their correctness.

\vspace{5mm}

\noindent
{\large\bf  2. Neutral current in the large $m_t$ limit }
\vspace{4mm}

The interaction of the Z boson to quarks is described
(in the lowest order approximation in the weak coupling constant)
by adding to  the QCD Lagrangian an extra term of the form
$M_Z\left(\frac{G_F}{2\sqrt 2}\right)^{1/2}Z^\alpha J^0_\alpha $,
with
$
J^0_\al
=
\sum_i \ovl{\psi}_i\g_\al(g^V_i -   g^A_i\g_5)\psi_i \
$
being the neutral  quark current.
\ice{
The upper
scripts $N$ and $NS$ stand for singlet and non-singlet contributions
respectively
(the exact
definitions  will be given  below).
}
The axial  part of the neutral current will be
conveniently rewritten as a sum over quark weak
isodublets $(\psi_i,\psi_{i'})$
\beq
A^0_\alpha = \sum_{i=u,c,t}
\D^{i}_\alpha, \ \  \D^{i}_\alpha= A^i_\alpha - A^{i'}_\alpha
\EQN{2.1}
\eeq
with
$A^i_\alpha=\ovl\psi_i \g_5\g_\alpha \psi_i$.
As for $\gamma_5$ in dimensional regularization
we employ essentially the definition of
refs. ~\cite{Hooft72,BM77a,gamma5mu}
\ice{
(the $\ep$-tensor is  a genuinely four-dimensional
object and, thus, the indices $\alpha,\beta,\nu,\rho$ below
are restricted  to  $0,1,2$ and $3$)
}
\beq
A^i_\alpha = \ovl{\psi}_i\g_\alpha\g_5 \psi_i \equiv
\frac{\xi^A_5(\ac)}{6}\epsilon _{\alpha\beta\nu\rho}
\ovl{\psi_i}\g_\beta\g_\nu\g_\rho \psi_i.
\EQN{axial.c.def}
\eeq
The finite normalization factor $\xi_5^A = 1-  4{\Ac}/{3}
+O(\Ac^2)$ on the rhs of  \r{axial.c.def}
is necessary  \cite{Trueman79,Collins84,Larin91,Larin92,CK3}
for  the current  \r{2.1} to obey the usual
(non-anomalous) Ward identities which in turn are crucial
in  renormalizing the standard  model.

Taken alone, the  axial vector quark current $A_\al^i$
has  non-zero  anomalous dimension
\beq
\mu^2\frac{d}{\mu^2} A_\al^i = \g^{\psi} (a) A^\psi_\al
=
\left
       (\sum_{n=0}-\g^{\psi}_n a^{n+2}
\right)A^\psi_\al, \ \  A^\psi_\al \equiv \sum_i A^i_\al
\EQN{2.2}
\eeq
with \cite{Collins78,Larin92,CK3}
\beq
\g^\psi_0  =   \frac{3}{4} C_F T , \  \
\g^\psi_1  =   C_A C_F T  (
                  \frac{109}{96}
                 )
               - C_F^2 T  (
                   \frac{9}{16}
                  )
               + C_F n_f T^2  (
                  \frac{1}{24}
                  ).
\EQN{Gpsi}
\eeq
%   Gpsi =
%
%       + ca*cf*tr * (
%          - 109/96*as^3
%          )
%
%       + cf*tr * (
%          - 3/4*as^2
%          )
%
%       + cf*tr^2 * (
%          - 1/24*n_f*as^3
%          )
%
%       + cf^2*tr * (
%          + 9/16*as^3
%          );
Here
$\mu$ is the  normalization point, $a = a(\mu)$  and
\beq
\mu^2\frac{d}{d\mu^2} = \mu^2\frac{\partial}{\partial\mu^2}
+ \beta(\ac) \frac{\partial}{\partial \ac}
+ 2{m}^2 \gm(\ac)\frac{\partial}{\partial {m}^2}.
\EQN{2.3}
\eeq
$\dsp\beta(\ac)=\sum_{n=0}-\beta_n
a^{n+2}$ and
$\dsp\g_m(\ac)=\sum_{n=0}-\g^m_n
a^{n+1}$
are the QCD $\beta$-function and
the quark mass anomalous dimension  respectively.
$C_F$ and $C_A$ are the Casimir operators of the quark and the
adjoint representations of the color group while $T$ is
the trace normalization of the quark representation.
Due to the flavour independence of the anomalous
dimension $\g^\psi$ the currents $A^0$ and
$\D^i$ still remain scale-independent (that is
their anomalous  dimensions  vanish in every
order in $\al_s$) as they must be.

Dimensional regularization automatically  respects the vector Ward
identity and hence  no extra finite normalization factor
is needed in defining the vector part of the  neutral current:
\beq
V^0_\alpha = \sum_{i}
g_i^V V_\alpha^i , \  \  \ V_\alpha^i = \ovl{\psi}_i\g_\al \psi_i.
\EQN{vector}
\eeq
The functional form of the
light vector quark current remains untouched  after integrating out a
heavy quark and rewriting the  current in terms of effective (that is
properly normalized) light quark fields  (for a recent discussion
see e.g.
\cite{me93}).
Explicitly,
\beq
[V^i_\al]^{(6)}
\bbuildrel{=\!=\!\Longrightarrow}_{\scriptstyle{m_t\to\infty}}^{}
         [V^i_\al]^{(5)} + O(1/m_t)  \ \  \mbox{\rm if $i\not = t$},
\ \ \ \
[V^t_\al]^{(6)}
\bbuildrel{=\!=\!\Longrightarrow}_{\scriptstyle{m_t\to\infty}}^{}
        O(1/m_t).
\EQN{vector.eff}
\eeq
\ice{
\beq
[V^i_\al]^{(6)}
\bbuildrel{=\!=\!\Longrightarrow}_{\scriptstyle{m_t\to\infty}}^{}
\left\{
        \ba{ll}
         [V^i_\al]^{(5)} + O(1/m_t)  & \mbox{\rm if $i\not = t$}
         \\
         O(1/m_t)                & \mbox{\rm otherwise}
         \ea
\right.
\EQN{vector.eff}
\eeq
}
The  bracketed   $(6)$ and $(5)$ here and in the following
indicate that the running coupling constant, an operator  or a
function is to be taken  in the full QCD  with $n_f=6$ or in the
effective theory with $n'_f=n_f-1$  ``effective'' flavours
respectively.  By definition the asymptotic expansion
\r{vector.eff}  means that any Green function of light fields with an
insertion of, say, the  operator $[V^i_\al]^{(6)}$ can be approximated
by an "effective" Green function to be obtained
through replacing the heavy current by the rhs of \r{vector.eff}.
\ice{
The corresponding equations for the coefficient functions  are
usually called  matching equations.
}
These Green functions are to be calculated  by means of the effective QC
Lagrangian with the properly changed coupling constant and quark
masses \cite{Bernreuther,Marciano}.

The perverse nature of the axial vector current  manifests
itself when constructing the axial part of the effective
topless neutral current. It may not be obtained in  analogy to
\r{vector.eff} by just discarding the top quark contribution. This
would lead, according to \r{2.2}, to  a current
which would be  {\em non-invariant} with respect to  a
change of the renormalization scale  \cite{CK3} ---
a physically non-acceptable conclusion clearly contradicting the RG
invariance of the initial quark  current.

The problem was { \em in principle }completely solved in two recent
works \cite{CK3,CK4} by J.H. K\"uhn and one of the authors of the
present paper (K.G. Ch.) along the lines  outlined by
J. Collins, F. Wilczek and A. Zee in  \cite{Collins78}.
It has been derived  there that the correct version
of \r{vector.eff} for the axial vector case  reads:
\beq
[A^i_\al]^{(6)}
\bbuildrel{=\!=\!\Longrightarrow}_{\scriptstyle{\ovl{m}_t\to\infty}}^{}
[A^i_\al]^{(5)}
+
{\rm \bf C}_\psi (\ac_{6}(\mu),\mu/\ovl{m}_t)  [A^L_\al]^{(5)},
\ \ \
i \not = t,
\EQN{2.4}
\eeq
\beq
[A^t]^{(6)}
\bbuildrel{=\!=\!\Longrightarrow}_{\scriptstyle{\ovl{m}_t\to\infty}}^{}
{\rm \bf C}_h (\ac_{6}(\mu),\mu/\ovl{m}_t)     [A^L]^{(5)}
\EQN{2.5}
\eeq
with the coefficient functions \ice{(CF)}
${ \bf C}^\psi$ and
${ \bf C}^t$
to be obtained  from the corresponding matching equations and
$A^L_\al = \sum_{i=u,d,s,c,b}A^i_\al $.
Here  $\ovl{m}_t = \ovl{m}_t(\mu)$ is the running mass of the
top quark  in the  $\msbar$ scheme.

All the operators entering into expansions \r{2.4} and \r{2.5} are
assumed  to be minimally renormalized at the scale
$\mu=\mu_t \approx m_t$.
\ice{While computing the CF's ${\bf C_^\psi$ and ${\bf C_}}^t$
it is only natural }
It follows from \r{2.4}, \r{2.5} that the current $\D^i_\al$
with $i\not=t$ behaves  in the heavy top quark mass limit in
exactly the same way as the vector current. The expansion of
$\D^t_\al$ expressed in terms of the  effective light quark
operators normalized at a lower mass scale
$\mu \approx M_Z$ reads
(if not stated otherwise, all quantities
below are  to taken in the effective QCD with $n_f=5$;
for a detailed derivation of
\r{2.7} see ref. \cite{CK4})
\vspace{-10mm}

\beq
\ba{c}
\dsp
[\D^t_\al]^{(6)}  =  - \left[A^b_\al\right]\trpsbs{\mu}{}
\\
\dsp
+\left\{{\bf C}_h(\ac_{(6)}(\mu_t),\frac{\mu_t}{\ovl{m}_t(\mu_t)})
-{\bf C}_\psi(\ac_{(6)}(\mu_t),\frac{\mu_t}{\ovl{m}_t(\mu_t)})
-
\frac{\g^{\psi}_0}{\be_0}\left(\Ac(\mu_t) - \Ac(\mu )\right)
\right\}
\left[A^L_\al\right] \trpsbs{\mu}{}
\\
\dsp
-\left\{
\frac{5}{2}\left(\frac{\g^{\psi}_0}{\be_0}\right)^2
\left(\Ac(\mu_t) - \Ac(\mu )\right)^2
\left.
+\frac{1}{2}
\left(
\frac{\g^{\psi}_1}{\be_0}
-
\frac{\g^{\psi}_0\be_1}{\be^2_0}
\right)
\left(\Ac^2(\mu_t) - \Ac^2(\mu )\right)
\right\}
\left[ A^L_\alpha \right]\trpsbs{\mu}{}.
\right.
\ea
\EQN{2.7}
\eeq
The notation $[ O ]_\mu$ indicates that the operator $[O]$ is
renormalized at scale $\mu$.
In the rhs of \r{2.7} power
suppressed terms  as well all  terms of order $a^{n}(\mu )
\ln^{n'}(\mu_t^2/\mu ^2)$ with $n - n' \ge 3$ are neglected.
Furthermore  all higher  logarithmic terms such as
$\as^{1+n}(\ln(M_Z^2/m^2_t))^n$ and $\as^{2+n}(\ln(M^2_Z/m^2_t))^n$
are  resummed  \cite{CK4}.

\ice{
Note  that the appearance  of {\em two}
renormalization scales in
}

An important feature of the relation \r{2.7} is that
it contains no explicit logarithms of $M_Z^2/m_t^2$ and
thus displays  a smooth  limiting behaviour  as
$m_t \to \infty$.

We have calculated  the $\al^3_s$ corrections  to the
coefficients entering
${\rm \bf C}_t$  and ${\rm \bf C}_\psi$
using the so-called ``method of projectors"
(see \cite{Gor87} and references therein).
Within the method the problem is reduced to
evaluating some three loop massive integrals {\em without}
external momenta. The latter task was performed  using
a program   written in
FORM \cite{FORM}. The program  implements
an algorithm developed in \cite{Broadhurst92}, which in turn
is based on the methods  of ref.  \cite{me81a,me81b}.
Our results read (in both formulas below $a = a_{(6)}(\mu)$;
terms of order $a^2$ were first computed in \cite{Collins78,CK4})
\beq
\ba{c}
\renewcommand{\arraystretch}{3}
\dsp
   {\bf C}_h(a,\frac{\mu}{m_t})=
         a^2 C_F T
        \left(
            \frac{3}{16}
          - \frac{3}{4} \ln(\frac{\mu^2}{m_t^2})
        \right)
\\
\dsp
       + a^3 C_F  T
\left\{
         C_F
       \left(
           \frac{5}{32}
          - \frac{3}{2} \zeta_3
          + \frac{27}{16} \ln(\frac{\mu^2}{m_t^2})
              \right)
       +     n_f T   \left(
           \frac{473}{432}
          - \frac{1}{6} \ln(\frac{\mu^2}{m_t^2})
          + \frac{1}{4} \ln^2(\frac{\mu^2}{m_t^2})
          \right)
\right.
\\
\dsp
\left.
  + C_A
  \left(
          - \frac{4621}{1728}
          + \frac{29}{16} \zeta_3
          - \frac{19}{24} \ln(\frac{\mu^2}{m_t^2})
          - \frac{11}{16} \ln^2(\frac{\mu^2}{m_t^2})
          \right)
       -  \frac{3}{4} T

\right\},
\ea
\EQN{CFH}
\eeq
\beq
\dsp
   {\bf C}_\psi (a,\frac{\mu}{m_t}) =
        a^3  C_F T^2
       \left(
          - \frac{343}{432}
          - \frac{1}{24} \ln(\frac{\mu^2}{m_t^2})
          - \frac{1}{4} \ln^2(\frac{\mu^2}{m_t^2})
      \right).
\EQN{CFL}
\eeq

\newpage

\noindent
{\large\bf  3.  Two-point correlators in the large $m_t$ limit}
\vspace{4mm}

In studing the QCD corrections to the Z width it
is both customary and useful  to express the decay rates  $\G_Z^V$
and $\G_Z^A$  through  some two-point correlator functions:
\beq
%\G_Z^h = \G_Z^V + \G_Z^A,  \  \
\G_Z^V = \G^{QPM}\sum_{i,j} g_i^V g_j^V R^V_{i,j},
\ \  \ \ \ \ \ \
\G_Z^A  = \G^{QPM}\sum_{i,j} g_i^A g_j^A R^A_{i,j}.
\EQN{3.1}
\eeq
Here the spectral density    $R^{V/A}_{i,j}$ is  defined as
\beq
R^{V/A}_{i,j} = 2\pi i \lim_{\delta\to 0}
\left(
\Pi^{(1),{V/A}}_{i,j}(q^2 = s - i\delta)
-
\Pi^{(1),{V/A}}_{i,j}(q^2 = s +  i\delta)
\right)
\EQN{3.2}
\eeq
where $\Pi^{V/A}_{i,j}(q^2)$ is       of the transversal part
of the correlator
\beq
 \Pi^{V/A}_{\mu\nu,i,j}(q)= i \int dx e^{iqx}
\langle
T[j^{V/A}_{\mu,i}(x)j^{V/A}_{\nu,j}(0)^\dagger] \rangle,
\ \ J^{V/A}_{\mu,i}(x) =  \ovl{\psi}_i\g_\mu(\g_5) \psi_i
,
\EQN{correllator.def.1}
\eeq
\beq
 \Pi^{V/A}_{\mu\nu,i,j}(q)=
(-g_{\mu\nu}q^2+q_{\mu}q_{\nu}) \Pi^{(1),{V/A}}_{i,j}(q^2)
                       + q_{\mu}q_{\nu} \Pi^{(0),V/A}_{i,j}(q^2).
\EQN{correllator.def.2}
\eeq
\ice{
The ratio
\beq
R(s)
= \sum_{i,j}e_i e_j R^V_{i,j}(q^2 =s)
,
\EQN{R}
\eeq
with   the electromagnetic quark current
$J^{EM}_\al = \sum_i e_i J^V_{\al,i}$.
}

All diagrams contributing to the correlator \r{correllator.def.2} are
naturally divided into two classes:  {\em non-singlet} diagrams which
contain the both external current vertices inside a single fermion
circle, and the rest --- the singlet ones --- in which every
(external) current vertex is attached to  its own fermion loop.  The
non-diagonal correlator $\Pi^{V/A}_{i,j}$, $i\not=j$ receives
contributions from singlet diagrams only while the diagonal one
$\Pi^{V/A}_{i,i}$ includes both types of diagrams.  An important
feature of non-singlet diagrams is that they do not discriminate
between vector and axial vector cases. This is because the definition
\r{axial.c.def} does  effectively restore the naive anticommutativity
proprety of the $\g_5$ matrix \cite{Larin92,BroadKatraev93}.

As a result, one may decompose the spectral densitiy
$R^{V/A}_{i,j}$ as follows\footnote{
We consider $u,d,s,c$ and $b$  quark as massless ones;
higher order corrections to $\G_Z^h$  proportional to
$m_b^2$ were studied in
\cite{ChetKuhn90,ChetKuhnKwiat92,ChetKwiat92}.}
\beq
R^{V/A}_{i,j} = \delta_{ij} R^{NS} + R^{V/A,S}, \ \
\ \
R^{NS} = 1 + \sum_{i \geq 1}^{j < i}
d^{NS}_{i,j} a^i \ell^j ,
\ \
R^{V/A,S} = \sum_{i \geq 2}^{j < i}
d^{V/A,S}_{i,j} a^i\ell^j,
\EQN{3.5}
\eeq
with
$
\dsp
\ell = \ln\frac{\mu^2}{s}.
$
The first relation in the chain  \r{3.5} directly
leads\footnote{We explicitly use the fact that for a every
weak isodublet $(\psi_i,\psi_{i'})$ $g_i^A = -g_{i'}^A = 1$.}
to the following representations
of non-siglet ($\G_Z^{A,NS}$) and singlet ($\G_Z^{A,S}$)
parts of $\G_Z^A$:
\beq
\G_Z^{A,NS} = \G^{QPM} \cdot  5 R^{NS},
\ \
\G_Z^{A,S} = \G^{QPM}  \cdot  R^{\D,S}_{tt},
\ \ \
R^{\D,S}_{tt}  = R^{\D}_{tt} - R^{NS},
\EQN{3.6}
\eeq
with
$$R^{\D}_{t,t}= R^A_{t,t} - R^A_{t,b} -R^A_{b,t}+ R^A_{b,b}$$
being the spectral density for the correlator
of the currents $\D^t_\mu$ and $\D^t_\nu$.

Now, keeping in mind the asymptotic expansions
(\ref{vector.eff}-\ref{2.7}),
one immediately arrives at the conclusion that
all we need to construct $\G_Z^A$ and $\G_Z^V$
in the large $m_t$ limit
is the knowledge  of the respective (massless!)
correlators of vector and axial
vector currents in the effective QCD
with five active flavours only.

To  find  the $\al_s^3$ contribution to
$\G_Z^A$    the relations \r{CFH} and  \r{CFL} are required
and, furthermore, the coefficients
$
d^{A,S}_{3,j}, \ \ j = 0,1,2.
$
The values of  $d^{A,S}_{2,j}$ and
$d^{A,S}_{3,j+1}$ with $j=0,1$ have been found
in \cite{CK4} and read\footnote{Unfortunately, the corresponding
eqs. (22) in \cite{CK4} are erroneously missing a factor 2 in
front of terms proportional to $\beta_0$. The error led to
numerically wrong coefficients in eqs. (24) and ((27) of
\cite{CK4}. Yet, the main conclusions of ref. \cite{CK4}
as well as its main results as expressed by eqs. (25), (26)
and (28) remain unchanged.}
\beq
d^{A,S}_{2,0} = -\frac{17}{12} C_A C_F T,
%d^{A,S}_{2,0} = -17/6,
\ \
d^{A,S}_{2,1} = -2\g^\psi_0,
\ \
d^{A,S}_{3,2} = \beta_0 d^{A,S}_{2,1},
\ \
d^{A,S}_{3,1} = - 2 (\g^\psi_0  + \g^\psi_1) + 2 \beta_0 d^{A,S}_{2,0}.
\EQN{3.7}
\eeq
We have performed the calculation of $d^{A,S}_{3,0}$  using
essentially the same methods as those used in \cite{Gorishny1} (for a
recent review see \cite{AIP93}).  These  include the IR rearrangement
\cite{Vladimirov78,me80}, the  $R^*$-operation \cite{me84}
and the integration by parts method \cite{me81a,me81b} as well a compute
implementation of the latter \cite{mincer2} on the basis of the
algebraic manipulation language  FORM \cite{Ver91}.  Our result
for $d^{A,S}_{3,0}$ is ($n'_f = n_f -1 =5 $)
\beq
\ba{c}
\dsp
d^{A,S}_{3,0} =
\\
\dsp
C_A C_F T
\left\{
        n'_f T        \left(
            \frac{923}{648}
          - \frac{1}{18} \pi^2
                       \right)
       + C_F     \left(
           \frac{1}{2} \zeta(3)
          - \frac{53}{96}
                 \right)
       + C_A      \left(
           \frac{11}{72} \pi^2
          - \frac{16231}{2592}
          + \frac{17}{24} \zeta(3)
                   \right)
\right\}.
\ea
\EQN{3.8}
\eeq
In the course of our calculation we also
found the spectral density of the longitudinal
(that  is corresponding to the spin zero)
part of the correlator \r{correllator.def.1}.
It is obviously nonvanishing only for the axial
vector currents and reads ($a_{(5)} = a_{(5)}(\mu)$)
\beq
R^{(0),A}(s) =
a_{(5)}^2 \, C_A C_F T
\left\{
           \frac{1}{4}
       -     a \, n'_f T     \left(
            \frac{1}{6} \ln(\frac{\mu^2}{s})
          + \frac{7}{12}
                             \right)
       +  a_{(5)} \, C_A    \left(
           \frac{97}{48}
          + \frac{11}{24} \ln(\frac{\mu^2}{s})
                      \right)
\right\}.
\EQN{3.9}
\eeq
Note that  the $R^{(0),A}$ will in general contibute
to the decay rate of the off-shell
$Z$ boson\footnote{K.G.Ch. thanks V.A. Ilyin and
E.E. Boos for a useful  discussion of the point.}.
\vspace{5mm}

\noindent
{\large\bf 4.  Correction of order $\alpha^3_s$ to the Z decay rate}
\vspace{4mm}

In the vector case the coefficients  appearing in \r{3.5}
are known up to and including terms of order $\al_s^3$
\cite{Sapirstein,Gorishny1}. Using \r{vector.eff}  one
 obtains
\beq
\ba{c}
\dsp
\frac{\G^V_Z}{\G^{QPM} } =
\left[
\sum_{i\not=t}  (g_i^V)^2
\left(
1 +  a_{(5)} + a^2_{(5)}  d_{2,0}^{NS} + a^3_{(5)}  d_{3,0}^{NS}
\right)
+
(\sum_{i\not=t} g_i^V)^2
\left(
a^3_{(5)} d_3^{V,S}
\right)
\right],
\\
a_{(5)} = a_{(5)}(M_Z), \ \
d_{2,0}^{NS} = 1.40923,\ \  d_{3,0}^{NS}=-12.76706, \ \
d_{3,0}^{V,S} = -0.41318 .
\ea
\EQN{G.vector.3}
\eeq
According to \r{3.6} the non-singlet part of $\G_Z^A$  is also easily
expressed in terms of  $d_{2,0}^{NS}$ and  $d_{3,0}^{NS}$, namely
\beq
\dsp
\G^{A,NS}_Z = \G^{QPM}
\left[
5
\left(
1 +  a_{(5)}(M_Z) + a^2_{(5)}(M_Z)  d_{2,0}^{NS}
+ a^3_{(5)}(M_Z)  d_{3,0}^{NS}
\right)
\right].
\EQN{G.axial.NS.3}
\eeq

At this point one  may proceed in two  ways.
First, one could plainly use \r{CFH} and  \r{CFL} as well
\r{3.6} and \r{3.7} to derive
\[
\ba{c}
\dsp
\G^{A,S}_Z = \G^{QPM}
\left[ f_2 a^2_{(5)}(\mu)
+ f_3  a^3_{(5)}(\mu)
\right],
\ea
\]
\beq
\ba{l}
\dsp
  f_2   =
        C_A C_F T    \left(
          - \frac{37}{24}
          + \frac{1}{2} \ln(\frac{M_Z^2}{\ovl{m}_t^2})
                    \right),
\\
\dsp
    f_3  =
           C_A C_F n'_f T ^2 \left(
          + \frac{25}{36}
          + \frac{11}{12} \ln(\frac{\mu^2}{M_Z^2})
          + \frac{1}{6} \ln^2(\frac{\mu^2}{M_Z^2})
          - \frac{1}{18} \pi^2
          + \frac{1}{9} \ln(\frac{\mu^2}{\ovl{m}_t^2})
          - \frac{1}{6} \ln^2(\frac{\mu^2}{\ovl{m}_t^2})
          \right)
\\
\dsp
       + C_A^2 C_F T    \left(
          - \frac{215}{48}
          - \frac{1}{2} \zeta_3
          - \frac{161}{48} \ln(\frac{\mu^2}{M_Z^2})
          - \frac{11}{24} \ln^2(\frac{\mu^2}{M_Z^2})
          + \frac{11}{72} \pi^2
          + \frac{19}{36} \ln(\frac{\mu^2}{\ovl{m}_t^2})
          + \frac{11}{24} \ln^2(\frac{\mu^2}{\ovl{m}_t^2})
          \right)
\\
\dsp
       + C_A C_F^2 T  \left(
          - \frac{3}{4}
          + \frac{3}{2} \zeta_3
          - \frac{3}{4} \ln(\frac{\mu^2}{\ovl{m}_t^2})
          \right)
       + C_A C_F T ^2  \left(
          - \frac{41}{54}
                         \right),

\ea
\EQN{4.1}
\eeq
where we have utilized
the  equality
$a_{(6)}(\mu) = a_{(5)}(\mu) +
a^2_{(5)}(\mu) \frac{T}{3} \ln\frac{\mu^2}{\ovl{m}_t^2}
+ O(a^3)
$ to express   $a_{(6)}(\mu)$ appearing in
\r{CFH} and  \r{CFL} through $a_{(5)}(\mu)$
(see e.g.  \cite{Bernreuther}).
Another possibility is to use
the relation \r{2.7}  ``RG improved" via summing
higher logs     with the result
\cite{CK4}
\beq
\ba{c}
\dsp
\frac{\G^{A,S}_Z}{\G^{QPM}} =
\dsp
 2 \frac{\g^{\psi}_0}{\be_0}
(a(\mu_t) - a(\mu_Z))R^A_{L,b}
 - 2
\left(
{\bf C}_h(\ac_{(6)}(m_t),1)
-{\bf C}_\psi(\ac_{(6)}(m_t),1)
\right)
R^A_{L,b}
+R^{A,S}_{b,b}
\\
\dsp
+ \left(
\frac{\g^{\psi}_1\be_0 - \g^{\psi}_0\be_1}{\be^2_0}
\right)
\left(\Ac^2(\mu_t) - \Ac^2(\mu_Z)\right)R^A_{L,b}
\ice{
\\
\dsp
}
+ \left(\frac{\g^{\psi}_0}{\be_0}\right)^2
\left(\Ac(\mu_t) - \Ac(\mu_Z)\right)^2
\left(5 R^A_{L,b} +  R^A_{L,L}\right),
\ea
\EQN{4.3}
\eeq
where  $\dsp R^A_{L,b}= \sum_{i=u,d,s,c,b} R^A_{i,b}$
and $\dsp R^A_{L,L} = \sum_{i,j=u,d,s,c,b} R^A_{i,j}$.
All the spectral densities in the rhs of eq. \r{4.3} are  to
be evaluated at $s = M_Z^2$ and the normalization scale $\mu$.
In deriving \r{4.3} we have used \r{2.7} with $\mu_t = m_t$.
The scale-invariant mass  $m_t$ is defined to be
the solution of the equation $\ovl{m}_t (\mu = m_t) = m_t$.

Explicitly,  \r{4.3} reads (from now on we  set $C_F =4/3,
\ \ C_A = 3$ and $T = 1/2$)
\beq
\ba{l}
\dsp
\frac{\G^{A,S}_Z}{\G^{QPM}} =
 \frac{12}{23}\left(a_{(5)}(m_t)-a_{(5)}(\mu)\right) (1 + a_{(5)}(\mu))
\\
\dsp
          - \frac{1}{4} \left(1 + a_{(5)}(\mu)\right)
             a_{(6)}^2(m_t)
       + a_{(6)}^3(m_t)    \left(
           \frac{7789}{1296}
          - \frac{55}{12}\zeta_3
          \right)
\\
\dsp
       + a_{(5)}^2(\mu)    \left(
          - \frac{17}{6}
          - \ln(\frac{\mu^2}{s})
          \right)
       + a_{(5)}^3(\mu)    \left(
          - \frac{41371}{1296}
          + \frac{67}{12}\zeta_3
          - \frac{373}{24}\ln(\frac{\mu^2}{s})
          - \frac{23}{12}\ln^2(\frac{\mu^2}{s})
          + \frac{23}{36}\pi^2
          \right)
\\
\dsp
+ \frac{360}{529}\left(a_{(5)}(m_t)-a_{(5)}(\mu)\right)^2
+ \frac{4007}{6348}\left(a_{(5)}^2 (m_t)  -a_{(5)}^2(\mu)\right).
\ea
\EQN{4.4}
\eeq
The difference between \r{4.1} and \r{4.3} originates from
the different ways of taking into account higher logs of the
form
$\as^{1+n}(\ln(M_Z^2/m^2_t))^n$ and $\as^{2+n}(\ln(M^2_Z/m^2_t))^n$.
In \r{4.3} they all are summed up (provided, of course,
the  normalization scale  $\mu_Z$
is chosen to be of the order    $M_Z$)
while in \r{4.1} they  are neglected
\cite{CK4}. Numerically the difference is negligible for any
reasonable value of $m_t$ \cite{CK4}.
However, the eq. \r{4.3} should be better
suited to study the Z boson propagator at energies
$\sqrt{s} << M_Z  $.

\vspace{5mm}

\noindent
{\large\bf 5.  Discussion}
\vspace{5mm}

An important feature of the separation of
$\G_Z^A$ into singlet and non-singlet parts
is that  both pieces individually must be
renormalization group invariant \cite{CK4}.
It was found in \cite{CK4} that (i) the lowest order result
\r{Kuhn1}  has rather strong dependence on $\mu$ when
$\mu$ is varied between $M_Z/2$ and $2 M_Z$ and that
(ii) the $\mu$ dependence gets much less pronounced
after including the logarithmic terms of order
of $\as^3$. Thus we may fix below $\mu = M_Z$.

Expressed through  the scale invariant mass $m_t$,
$\G_Z^{A,S}$   assumes the form
(the terms proportional to $\ln(\frac{m_t^2}{M_Z^2})$ and
$\ln^2(\frac{m_t^2}{M_Z^2})$
on the second line
were first found in \cite{CK4})
\beq
\ba{c}
\G^{A,S}_Z  =  \G^{QPM}\times
\dsp
\left\{
\left(
- \frac{37}{12} - \ln(\frac{m_t^2}{M_Z^2})
\right) a_{(5)}^2(M_Z) +
\right.
\\
\dsp
\left.
\left(
- \frac{5651}{216} + \zeta(3) + \frac{23}{36}\pi^2
+ \frac{23}{12}\ln(\frac{m_t^2}{M_Z^2})^2
- \frac{67}{18}\ln(\frac{m_t^2}{M_Z^2})
\right)a_{(5)}^3(M_Z)
\right\}
\ea
\EQN{5.1}
\eeq
or, numerically,
\beq
\ba{c}
\G^{A,S}_Z  =  \G^{QPM}\times
\dsp
\left\{
\left(
- \frac{37}{12} - \ln(\frac{m_t^2}{M_Z^2})
\right) a_{(5)}^2(M_Z) +
\right.
\\
\dsp
\left.
\left(
-18.65440
+ \frac{23}{12}\ln^2(\frac{m_t^2}{M_Z^2})
- \frac{67}{18}\ln(\frac{m_t^2}{M_Z^2})
\right)a_{(5)}^3(M_Z)
\right\}.
\ea
\EQN{5.1b}
\eeq
Note that in agreement with the estimation
of \cite{CK4}  the correction of order $\alpha_s^3$
amounts to less than 30\% change of the lowest order
result for $\G^{A,S}_Z$ even for $\alpha_s(M_Z)$ as large
as $0.13$ and $m_t$ as low as $M_Z$.

At last, we write our  final result for total hadronic decay
rate of the $Z$ boson
\beq
\ba{c}
\dsp
\G_Z=\frac{G_F M^3_Z}{8\pi\sqrt2}
\left[
\sum_{i=u,d,s,c,b}  (g_i^V)^2
\left(
1 +  a_{(5)} + 1.40923 a^2_{(5)} -  12.76706 a^3_{(5)}
\right)
\right.
\\
\dsp
\left.
+
(\sum_{i=u,d,s,c,b} g_i^V)^2
\left(
-  0.41318 a^3_{(5)}
\right)
\right.
\\
\left.
\dsp
+\sum_{i=u,d,s,c,b}  (g_i^A)^2
\left(
1 +  a_{(5)} + 1.40923 a^2_{(5)} -   12.76706 a^3_{(5)}
\right)
\right.
\\
\left.
\dsp
+ \left(
- \frac{37}{12} - \ln(\frac{m_t^2}{M_Z^2})
\right) a_{(5)}^2 +
\left(
-18.65440
+ \frac{23}{12}\ln^2(\frac{m_t^2}{M_Z^2})
- \frac{67}{18}\ln(\frac{m_t^2}{M_Z^2})
\right)a_{(5)}^3
\right],
\ea
\EQN{5.2}
\eeq
where $ a_{(5)} = a_{(5)}(M_Z)$  and
$g_q^V = 2I_q -4Q_q s^2_W$,
$g_q^A= 2I_q$ for a quark  $q$.

After the main bulk of our calculations and various cross-checks was
finished we received   a preprint \cite{Larin93} on a related subject. I
this paper  the $O(\alpha_s^3)$  corrections to $\G_Z^A$ have been
computed by directly expanding  the relevant diagrams in the limit of
a large mass of the top quark.  The net result obtained in \cite{Larin93
amounts to eq. \r{4.1} but with  a bit different coefficients
of the both terms proportional to $\zeta(3)$. The disagreement
forced us to spend an extra  month carefully  testing our results.
Having failed to find an error in our calculations we let the authors
of  \cite{Larin93} know about the disagreement and within days they
discovered  a mistake in theirs. The corrected result of the authors
of \cite{Larin93} is now in the complete agreement with \r{4.1}.
\vspace{5mm}

{\large \bf 6.  Acknowledgments}
\vspace{5mm}

K.G.Ch. thanks  J.K.K\"uhn for a lot of very useful discussions and
valuable advice as well as  the permanent  support and encouragement
he has received  during all stages of  his work on  the project.  He
appreciates the warm hospitality of Institute of Theoretical Particle
Physics at the Karlsruhe University as well as  the  financial
support from BMFT and HERAEUS-Stiftung.  O.V.T. is grateful to the
Physics Department of Bielefeld University for warm hospitality, and
to BMFT for financial support.
We thank  S.A.  Larin  for e-mailing
us  the text of preprint \cite{Larin93} before its publication.  We
are  indebted  to Mrs. M. Frasure and   A.  Kwiatkowski for their
help in preparing the manuscript.
The work was partially supported by the  Russian
Fund of the Fundamental Research, Grant N 93-02-14428.

*\renewcommand{\baselinestretch}{0.8}
%\begin{verbatim}

%\end{verbatim}
\end{document}